\documentclass{sigchi}
\toappear{Permission to make digital or hard copies of all or part of this work for personal or classroom use is granted without fee provided that copies are not made or distributed for profit or commercial advantage and that copies bear this notice and the full citation on the first page. Copyrights for components of this work owned by others than the author(s) must be honored. Abstracting with credit is permitted. To copy otherwise, or republish, to post on servers or to redistribute to lists, requires prior specific permission and/or a fee. Request permissions from Permissions@acm.org.\\
{\emph{CHI 2017}}, May 06--11, 2017, Denver, CO, USA\\
Copyright is held by the owner/author(s). Publication rights licensed to ACM.\\
ACM 978-1-4503-4655-9/17/05\ldots\$15.00\\
DOI: http://dx.doi.org/10.1145/3025453.3025575}

\toappear{\copyright~Scott A.~Hale and Irene Eleta 2017. This is the author's version of the work. It is posted here for your personal use. Not for redistribution. The definitive version was published in CHI 2017, http://dx.doi.org/10.1145/10.1145/3025453.3025575}

\usepackage{balance}
\usepackage{graphics}

\usepackage{txfonts}
\usepackage{times}
\usepackage[pdftex]{hyperref}

\usepackage{color}
\usepackage{textcomp}
\usepackage{booktabs}

\makeatletter
\def\url@leostyle{%
  \@ifundefined{selectfont}{\def\UrlFont{\sf}}{\def\UrlFont{\small\bf\ttfamily}}}
\makeatother
\def\pprw{8.5in}
\def\pprh{11in}

\definecolor{linkColor}{RGB}{6,125,233}
\hypersetup{%
  pdftitle={Foreign-language Reviews: Help or Hindrance?},
  pdfauthor={Scott A. Hale and Irene Eleta},
  pdfkeywords={experiment; user-generated content; product reviews; e-commerce; multilingualism; bilingualism; internationalization and localization},
  bookmarksnumbered,
  pdfstartview={FitH},
  colorlinks,
  citecolor=black,
  filecolor=black,
  linkcolor=black,
  urlcolor=linkColor,
  breaklinks=true,
}

\usepackage{hypothesis}
\begin{document}

\title{Foreign-language Reviews: Help or Hindrance?}

\numberofauthors{2}
 \author{%
  \alignauthor{Scott A.~Hale\\
    \affaddr{Oxford Internet Institute, University of Oxford}\\
    \affaddr{The Alan Turing Institute}\\
    \email{scott.hale@oii.ox.ac.uk}}\\
  \alignauthor{Irene Eleta\\
    \affaddr{Barcelona Institute for Global Health (ISGlobal)}\\
    \affaddr{CIBERESP and Universitat Pompeu Fabra}\\
    \email{irene.eleta@isglobal.org}}\\
}

\maketitle

\begin{abstract}
The number and quality of user reviews greatly affects consumer purchasing decisions. While reviews in all languages are increasing, it is still often the case (especially for non-English speakers) that there are only a few reviews in a person's first language. Using an online experiment, we examine the value that potential purchasers receive from interfaces showing additional reviews in a second language. The results paint a complicated picture with both positive and negative reactions to the inclusion of foreign-language reviews. Roughly 26--28\% of subjects clicked to see translations of the foreign-language content when given the opportunity, and those who did so were more likely to select the product with foreign-language reviews than those who did not. 
\end{abstract}


\category{H.5.m.}{Information Interfaces and Presentation
  (e.g. HCI)}{Miscellaneous}

\section[Introduction]{INTRODUCTION}

The amount of content online in different languages is greatly increasing, and the early days of English-language dominance on the Web have given way to language pluralism online. On many large user-generated content platforms, less than half the content is in English and many users do not speak English as a native language. As Internet-penetration rates are already high in many English-speaking countries, future user growth (and the content contributed by these users) will be predominantly in non-English languages \cite{graham-inet-penetration}.

The multilingual Internet we have today fractures users and content across language divides. This is especially apparent on user-generated content and social media platforms where the content is contributed by end users. The majority of articles on Wikipedia, for example, exist in only one language edition \cite{hecht2010}.  Although the English edition of Wikipedia is the largest edition of the encyclopedia, there is still a large amount of information not available in English that is available in other language editions. Users of smaller-sized language editions experience the inequality in information across languages even more acutely.

Most research in cross-language information retrieval has focused on helping individuals locate information in different languages and the performance of machine translation has greatly increased in recent years, but large questions remain about when and how to present content in a language different from the presumed native language of the user (henceforth ``foreign-language content'') in user-generated content platforms. 
Within this article we consider the context of online user reviews and test the impact of foreign-language reviews on the perceived helpfulness of all reviews as a whole.

User product reviews are an excellent context in which to test the perceived usefulness of foreign-language content given the high importance that consumers place on the number and quality of reviews \cite{park2007} and the well-established link between user reviews and purchasing decisions \cite{ye2009}.
The findings have practical relevance not only for how travel, restaurant, shopping, and other websites present foreign-language reviews to their users, but also for the general acceptability of foreign-language content and translations on social media and other user-generated content platforms. This is especially relevant when user reviews or user-generated content in general is not available in the preferred language of a user.

\section[Hypotheses]{HYPOTHESES}
Using an experimental approach, we first perform a within-subjects comparison to examine the effects of including foreign-language reviews. We then perform a between-subjects comparison to understand how adding a user-interface affordance to see translations of foreign-language reviews moderates this effect. Finally, we run a further iteration of the experiment in which reviews are ``pretranslated'' (i.e., the user does not need to click to see the translation).

Based on the literature examined below we expect that more reviews are more helpful, but it is not clear how helpful the inclusion of foreign-language reviews is (even with the option to see translations of the reviews). 
Indeed, one possibility is that foreign-language reviews could confuse or distract users and thus lower the overall usefulness that consumers receive from all the product reviews. Under this thinking it would be better to hide all foreign-language reviews from users even though this means fewer overall reviews are available.

Alternatively, we may find that foreign-language reviews have a positive, helpful effect (particularly when the number of native-language reviews is low) given the findings that more reviews are generally received more positively. Even without translations, users may still be able to pick out a few words or cognates and derive useful information from the number of stars given by each reviewer. This leads to our first hypothesis that 
\hypothesis{h:number}{experimental conditions with more reviews will be rated more highly.}

While to the best of our knowledge no research has looked specifically at the helpfulness of foreign-language reviews, previous studies have established that the readability of reviews correlates with how helpful the reviews are perceived \cite{korfiatis2012}. This implies that 
\hypothesis{h:first}{where the number of reviews is the same between two conditions, the condition with more reviews in the user's first language will be rated more highly.}

Additional factors that increase the ability of consumers to understand the reviews should also further increase the perceived helpfulness. We specifically test the addition of an interface affordance to see translations of the reviews testing the hypothesis that 
\hypothesis{h:machinetrans}{the option to see translations of foreign-language reviews will increase the overall usefulness that users get from the reviews.}

\section[Background and case selection]{BACKGROUND AND CASE SELECTION}
Online user reviews affect product sales, but many factors contribute to the influence that reviews have on the decision-making processes of consumers. Example factors include the trust associated with reviewers, the quantity of reviews, the quantity of negative reviews, and the timing of reviews \cite{hu2008} as well as the quality of reviews \cite{korfiatis2012}. The quantity of reviews needed for making a decision about a product likely varies from context to context, but has not been well studied. We follow the decision of Park et al.~\cite{park2007} to use six reviews in the high-review condition of their experiment based on the results of a small focus group.

Despite the fact that many platforms have reviews and other content in multiple languages, research is lacking on how to best present reviews in multiple languages (if at all) and many different approaches are currently in use on major websites. The travel review website TripAdvisor, for example, generally shows reviews in reverse chronological order (most recent reviews first), but demotes foreign-language reviews so that they appear after all reviews in the language of the locale selected by the user. Google Play, a mobile app store, in contrast, hides foreign-language reviews making them completely inaccessible (although reviews from all languages are used to calculate the average rating of an app). Twitter, Facebook, and Google Plus all provide the option to see machine translations of foreign-language posts, and Facebook has experimented with showing machine translations in place of foreign-language posts.

\subsection{Case selection}
In order to understand the effects of displaying foreign-language reviews to users, we designed an online experiment in which subjects compared three London bicycle tours. Tours are the third largest category of attractions in London on TripAdvisor behind Nightlife and Shopping \cite{hale2016tripadvisor}, and they are also a good example of an \textit{experience good} \cite{nelson1970} where the quality and utility of the product can only be determined upon consuming it (i.e., upon taking the tour), and user reviews are especially important in such cases \cite{klein1998}.

As a large, international city with tourists from all over the world, London is an appropriate setting. Non-English content has grown quickly online and now accounts for a sizable portion of user-generated content online \cite{hale2016tripadvisor,pimienta}. On TripAdvisor 25\% of all reviews for London attractions are not in English, and 6\% of tourist attractions in London have more non-English than English-language reviews~\cite{hale2016tripadvisor}.

We specifically compare the effects of showing Spanish-language reviews to supplement a smaller number of English-language reviews. The selection of languages was driven by three principal reasons. First, prior work has shown a reasonable correlation between Spanish and English language reviews \cite{hale2016tripadvisor}. Second, Spanish is the second most spoken language in the United States where our subject pool is based and understanding the best way to handle Spanish and English reviews is important as an increasing amount of commerce in the United States is conducted in Spanish. Third, and most importantly, speakers of smaller-sized languages are more willing (or, perhaps more forced) to engage with foreign-language content~\cite{hale2014wiki}, and thus, the helpfulness that English-speakers derive from foreign-language reviews can be seen as a lower bound. We should expect speakers of other languages to derive as much or more helpfulness from foreign-language content.

\section[Study I]{STUDY I}
\subsection{Method and approach}

\begin{figure}
\includegraphics[width=\columnwidth]{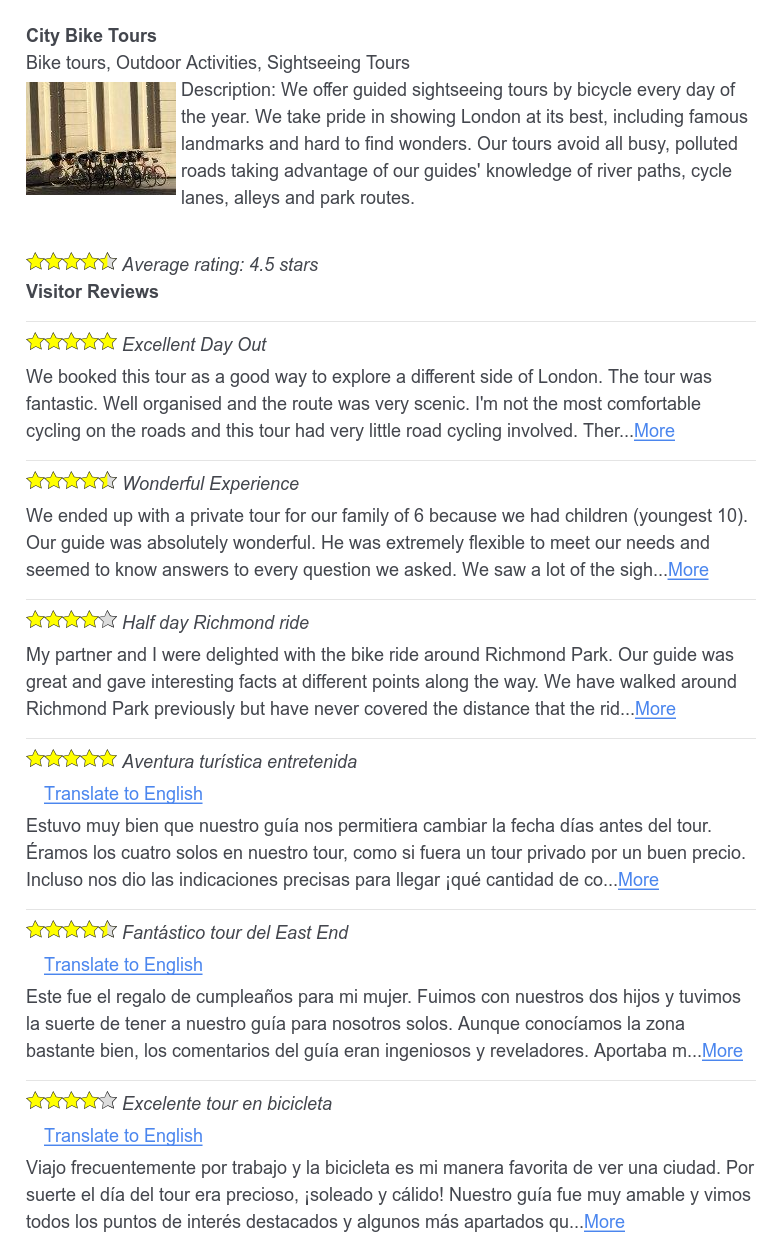}
\caption{The en3es3 condition (3 English reviews and 3 Spanish reviews) with the option to view a translation of each Spanish review. There was also a en3es3 condition with no option to view translations as well as a en3 condition with 3 English reviews and no Spanish reviews and a en6 condition with 6 English reviews and no Spanish reviews.}
\label{fig:expr_en3es3_1}
\end{figure}

In our first experiment, subjects were presented with three bicycle tours in London. The experimental interface showed a name, description, small photo, and three or six user reviews for each tour. There were a total of three different names, photos, and descriptions that were randomly grouped together for each subject and shown in a random order.\footnote{The exact wording of all content, the code used to run the experiment, and the anonymous data from the experiment itself are available at \url{http://scott.hale.us/pubs/?chi2017}.}

In our first iterations of the experiment, we recruited subjects using Amazon Mechanical Turk. We described the study as being about evaluating tourist attractions, and presented it similarly to the search engine optimization (SEO) tasks that are very common on Mechanical Turk. We are aware that Mechanical Turk users are not representative of the entire US population, but they are more diverse than traditional subject pools and have been used to replicate well-established experimental findings \cite{berinsky2012,buhrmester2011,goodman2013,mason2012,reips2000}. Furthermore, appropriate randomization procedures ensure subjects within the treatment and control conditions have similar demographics---something that is very different from annotating ``gold-standard'' datasets~\cite{sen2015}. 
This research constitutes a first step in understanding the potential effects that the inclusion of foreign-language reviews has on the user experience of a site, and 
we hope publication of this study will generate interest and allow us to work with a platform operator to test our findings in the wild.

To form our pool of reviews, we selected 15 reviews at random from London bicycle tours on TripAdvisor and removed tour-specific references such as the names of tour guides. The reviews on TripAdvisor of bicycle tours were overwhelmingly positive, and our pool consisted of only 4 and 5 star reviews. We (human) translated all reviews to Spanish to create a pool of 15 reviews with text in both Spanish and English.%
\footnote{One author is a native Spanish speaker and the other author is highly proficient.}
We assigned the reviews to each tour \emph{for each subject at random} such that we showed one tour with 3 English reviews (the \emph{en3} condition), one tour with 3 English and 3 Spanish reviews (the \emph{en3es3} condition), and one tour with 6 English reviews (the \emph{en6} condition). Within the en3es3 condition, half of the subjects were given the option to click to see English translations of the Spanish reviews and half were not. The en3es3 condition with the option to see English translations is shown in Figure~\ref{fig:expr_en3es3_1}.

Each review consisted of a star rating (1--5 stars), a title, and review text. We manipulated the star ratings so that each tour had a mean 4.5 rating overall (i.e., the tour in the en3 condition had one 5-star, one 4.5-star, and one 4-star review while the tours in the other two conditions had two reviews with each of these ratings). 
Review length has previously been shown to influence consumer decisions \cite{korfiatis2012}; so, we controlled for this by truncating all reviews at 250 characters and appending a link reading ``\dots More'' to read the review further. We did not present the names, locations, or other information about the authors of the reviews as such information has also been shown to affect the trustworthiness that users ascribe to reviews \cite{hu2008}.

We measured three outcome variables: the proportion of time spent evaluating each tour, an individual rating of the likelihood of booking each tour, and, at the end of the experiment, the selection of the \emph{one} tour that each subject would be most likely to book.

The first outcome was measured without any action by the subject. We tracked the amount of time spent evaluating the three tours (ignoring time spent on the consent form and answering ending questions) and calculated the proportion of time given to each condition. We looked specifically at the proportion of time that subjects spent in each condition as we expected the distribution of raw times to be skewed and long tailed as some subjects would read faster and complete the study more quickly than others.

For each tour, we asked subjects a self-reported likelihood of booking the particular tour using a slider or visual analogue scale (VAS) with the end points labeled as ``very unlikely'' and ``very likely''. This VAS resulted in an integer value between 0 and 100 inclusive, which we treated as continuous following common statistical practices.
The value was not shown to subjects, who had to simply place themselves visually along the continuum. Such VASs are very common for pain, depression, mood, and other subjective measures~\cite{mccormack1988,reips2008} having started in psychology in the early 1920's~\cite{hayes1921,freyd1923} and become more widespread at the end of the 1960's~\cite{mccormack1988}. Numerous studies have shown they are as accurate and reliable as Likert scales but have better sensitivity~\cite{joyce1975}. In particular, VASs are better suited for repeated measurements than Likert scales~\cite{joyce1975}.
Along with the likelihood to book, we asked subjects for the importance of the tour title, photo, description, and user reviews in reaching their decision. These were measured using VASs with endpoints labeled ``very unimportant'' and ``very important.''

After the subjects had evaluated all three tours, we asked them to indicate which one of the three tours they would be most likely to book if they were to book only one. 
We further asked the importance of the title, photo, description, and reviews in reaching this decision.

Once subjects completed this overall selection, we informed subjects about our interest in the effects of foreign-language reviews and asked for their self-reported abilities to read English and Spanish.
The exact wording of all questions is given in the appendix.

Per our hypotheses, we predicted higher ratings for the en3es3 condition than the en3 (more reviews are more helpful, H\ref{h:number}), but that the en6 condition would be rated even higher than the en3es3 condition (when the number of reviews is the same, users will prefer more reviews in their first language, H\ref{h:first}). In our between-subjects comparisons, we expected that subjects given the option to see translations of the Spanish-language text would rate the en3es3 condition more highly than subjects not given the ability to see translations (H\ref{h:machinetrans}). 

We first conducted a pilot of the experiment with 110 subjects before running the main experiment. Free-text comments from subjects indicated that our initial names for the tours were too unique and influencing subjects' choices. We changed the names from ``Easy Peddle Bike Tours,'' ``London Bicycle Tours,'' and ``Vintage Cycle Tours'' in the pilot to ``City Bike Tours,'' ``London Bicycle Tours,'' and ``Capital Cycle Tours'' for the main experiment. We further decided to randomize the matching of tour names, tour descriptions, and tour photos for each subject. We checked all reported results using control variables for tour name, description, and photo as well as control variables for the order in which tours were presented. 
The pilot also identified an issue with Google Chrome automatically translating foreign-language content, which was addressed in the full experiment.

\subsection{Results}

We paid subjects \$0.50 USD for this 5--7 minute study. We had 533 subjects complete the study, but discarded 53 subjects who either answered attention check questions incorrectly (e.g., what type of tour is this?) or indicated that they suspected the study was about foreign-language content before we informed them. We required subjects to be US-based and have a 95\% or higher acceptance rate to participate. 

Most subjects indicated they did not read Spanish at all (N=283) or that they read Spanish at only a basic level (N=143). We avoided asking about language abilities in advance in order to avoid biasing our subjects as to the purpose of our study. As a result, we had too few subjects with intermediate (N=36), advanced (N=14), or native (N=4) Spanish skills to conduct robust analysis with these groups. We set aside these subjects as well as 10 non-native speakers of English and proceeded with only the 416 subjects who were native English speakers with no or basic Spanish skills.

\subsubsection{Outcomes}
\begin{figure*}
\includegraphics[width=\textwidth]{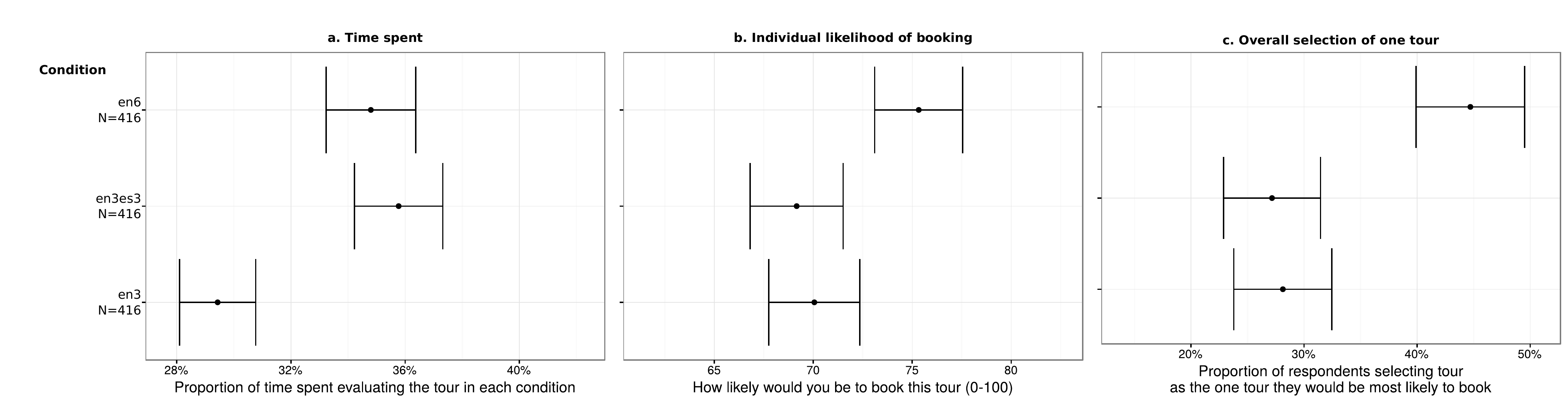}
\caption{Mean values with 95\% confidence intervals. (a) Subjects spent significantly more time in the en3es3 and en6 conditions than in the en3 condition. (b) Despite this additional time, subjects indicated they were more likely to book the tour shown in the en6 condition than tours in either of the other conditions and (c) were more likely to select this same tour when forced to choose the one tour they would most likely book.}
\label{fig:expr_outcomes}
\end{figure*}

\begin{figure*}
\includegraphics[width=\textwidth]{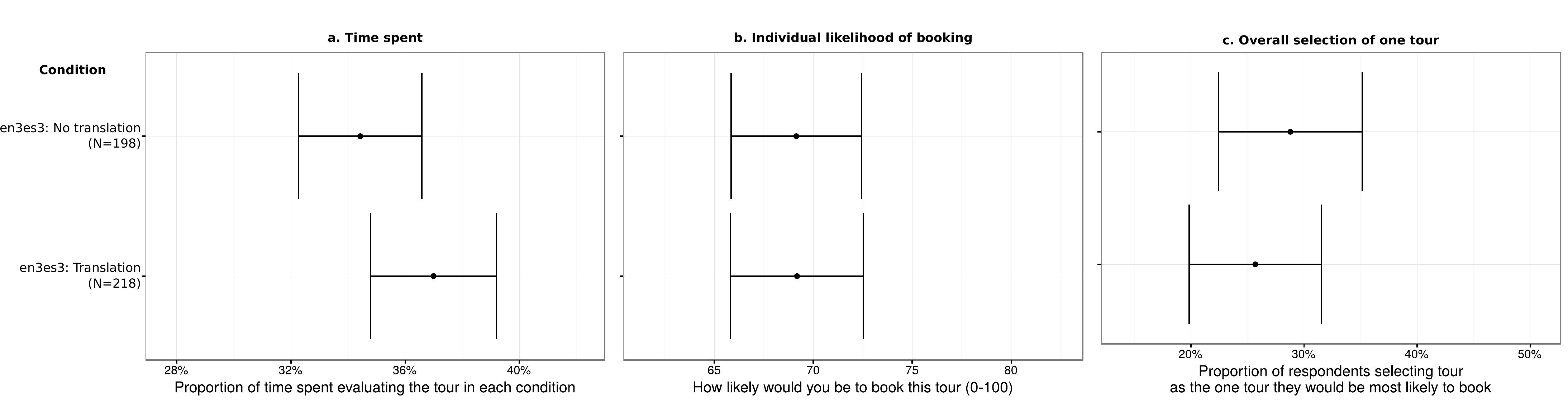}
\caption{Mean values with 95\% confidence intervals. The option to see translations of the Spanish language reviews was not associated with any statistical difference in (a) the amount of time, (b) the individual booking likelihoods, or (c) the overall selection of one tour.}
\label{fig:expr_outcomesE}
\end{figure*}

\begin{figure*}
\includegraphics[width=\textwidth]{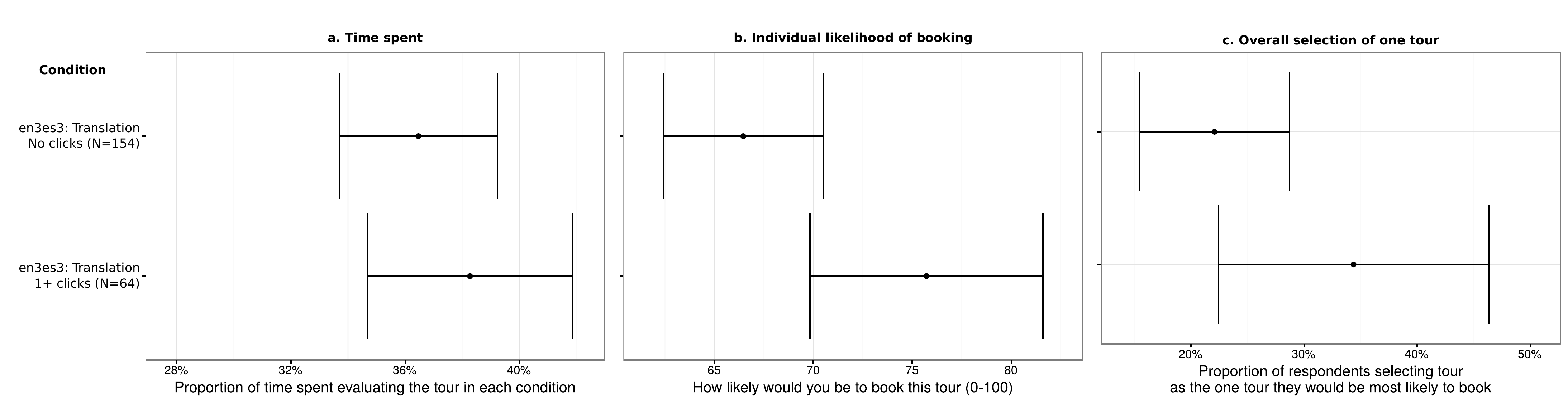}
\caption{Mean values with 95\% confidence intervals. 28\% of subjects clicked to see one or more translations in the en3es3 condition when given the option. Compared to subjects who did not click to see any translations the subjects who clicked to see translations (a) spent a similar amount of time, but (b) rated the tour more favorably and (c) were slightly more likely to select the en3es3 tour as the one overall tour they would most likely book.}
\label{fig:expr_outcomesE2}
\end{figure*}

Subjects on average spent four minutes (standard deviation 5.5) evaluating all three tours. In accordance with our first hypothesis, subjects spent a significantly smaller proportion of their time evaluating the tour in the condition with only three reviews (en3, 29\% of subjects' time on average) than either in the en3es3 condition (36\% of subjects' time) or the en6 condition (35\% of subjects' time), as shown in Figure~\ref{fig:expr_outcomes}a.
Two-tailed t-tests showed the difference between the en3 and en3es3 conditions as well as the difference between the en3 and en6 conditions was statistically significant ($p<0.003$ and $p<0.026$ respectively) while the difference between the en3es3 and en6 conditions was not ($p=0.38$). The same outcomes were produced with multivariate linear regressions that also controlled for the order, titles, photos, descriptions of the tours as well as the importance given to titles, photos, and descriptions.

In contrast to the amount of time spent evaluating each tour and contrary to our first hypothesis, when asked about their individual likelihood of booking each tour, subjects were no more likely to book the en3es3 tour than the en3 tour. The en6 tour was significantly more likely to be booked than either other condition (Figure~\ref{fig:expr_outcomes}b). Two-tailed t-tests produced p-values less than or equal to $0.001$ for both comparisons as did multivariate regressions with the control variables previously mentioned.

After evaluating all three tours (one in each condition), subjects were asked to choose the \emph{one} tour that they would be most likely to book. The results for this overall selection of one tour mirror those for the booking likelihood of each individual tour. 
Subjects were significantly more likely to select the tour presented with 6 English reviews (en6, selected by 45\% of subjects) than either other tour (Figure~\ref{fig:expr_outcomes}c).
28\% of subjects selected the en3 tour and 27\% selected the en3es3 tour. Both of these values were significantly lower than the 45\% selecting the en6 tour, established either with two-tailed t-tests ($p<0.001$ for both comparisons) or with multivariate logistic regression using the control variables previously mentioned.

These results only partially support our first hypothesis that the conditions with more reviews would be rated more favorably (H\ref{h:number}). While the en6 condition was rated most favorably in accordance with the hypothesis, we observed no difference between the individual or overall booking likelihoods of the tours in the en3 and en3es3 conditions despite the additional number of reviews in the en3es3 condition. At the same time, it is important to note that neither did we observe an overall negative effect created by the Spanish-language reviews as some interface designers could fear.

The differences between the en3es3 and en6 conditions strongly support the second hypothesis that when the number of reviews was the same, the condition with more reviews in subjects' first languages would be more highly rated (H\ref{h:first}).

\subsubsection{Translation}
Within the multilingual en3es3 condition, 218 subjects in the \textit{treatment} group were given the option to see English translations of the Spanish-language reviews while another 198 subjects in the \textit{control} group were shown the Spanish-language reviews with no option to see translations. Hypothesis H\ref{h:machinetrans} predicted that the subjects in the treatment group would rate the en3es3 tour more highly than the subjects in the control group.

Contrary to this hypothesis, there was no overall statistical difference between subjects in treatment and control. Subjects in the treatment group spent a slightly larger proportion of their time evaluating the tour in the en3es3 condition (37\% vs.\ 34\%, $p<0.03$ only when control variables were included), but were no more likely to book the tour.%
\footnote{There was no significant difference between control and treatment either in the individual ratings (69.1 vs.~69.2 points, $p=0.80$) or in the overall selection of the en3es3 tour as the most preferred (29\% vs.~26\%, $p=0.52$).}
Thus, the addition of the option to see translations of the Spanish-language reviews to the interface had no overall effect (Figure~\ref{fig:expr_outcomesE}).

These top-line figures are the result of two important underlying facts. First, most subjects ignored the option to see translations of the Spanish-language reviews: 72\% of the subjects in the treatment group did not click to see any translations. Among the 28\% of subjects who did click to see one or more translations, nearly all clicked to see all three translations (88\% of subjects who viewed one translation viewed all three translations).

Second, the subjects who chose to view one or more translations behaved very differently from the subjects who chose not to view any translations (Figure~\ref{fig:expr_outcomesE2}). Subjects clicking to view at least one translation rated the tour in the en3es3 condition significantly higher than subjects who did not click to view any translations. Subjects who clicked at least one translation rated the tour a mean 76 points while those not clicking a translation rated it a mean 66 points. This difference was statistically significant with both a two-tailed t-test ($p=0.01$) as well as a multivariate regression with the control variables ($p=0.002$).

The mean rating in the en3es3 condition for subjects clicking to see at least one translation (76 points) did not differ significantly from either their ratings in the en3 condition (mean 70, $p=0.08$) or in the en6 condition (mean 75, $p=0.90$). In contrast, the mean rating in the en3es3 condition for subjects not clicking any translation (66 points) was significantly lower than the en6 condition (mean 75, $p=0.0001$) and similar to their ratings in the en3 condition (mean 70, $p=0.13$).

The selection by subjects of one tour at the end of the experiment as the tour they would be most likely to book showed a similar pattern. Overall, 29\% of subjects in the control condition without translation buttons picked the en3es3 tour as the one they would be most likely to book. In contrast, only 26\% of subjects in the treatment condition picked the en3es3 tour. This again obscures differences in behavior between those who clicked or did not click to see translations. 22\% of the subjects in the treatment group who did not click to see any translations selected the en3es3 tour as the tour they would most likely book ($-7\%$ compared to control) while 34\% of subjects who clicked to see at least one translation selected the en3es3 tour as the tour they would be likely book ($+5\%$ compared to control).
The difference in the overall tour choice by subjects who clicked or not was significant when tested with logistic regression including the control variables previously mentioned ($p=0.02$) but was not significant with a t-test alone ($p=0.08$).
The opposite direction of these effects suggests adding the translation buttons (perhaps further drawing attention to the fact that some of the reviews were in a foreign-language) can actually have a negative effect among some people. The free-text comments made by subjects support the idea that the mere presence of foreign-language reviews can have an effect, a point we return to in the conclusions.

\subsection{Discussion}
Overall, adding Spanish language reviews had no effect on subjects' individual rating of tours nor on their overall selection of one tour. The addition of buttons to show translations of the Spanish reviews also had no overall effect.

However, subjects who used the translation buttons behaved very differently from those who did not use the translation buttons. The use of the translation buttons was very divisive: generally subjects either clicked none of the translation buttons or clicked all three translation buttons. Those who clicked rated the tour more highly and were more likely to select it overall compared to those who did not click any of the translation buttons.

We have found no published figures as to how much translation options are used in the wild (e.g., on Facebook, Twitter, TripAdvisor, etc.). Our uptake rate of 28\% is broadly in line with the figure an employee of one social media platform mentioned to us confidentially.

We did not hypothesize such a divergent effect for the Spanish-language content nor that the use of the translation buttons would distinguish subjects so clearly. As a result, we did not ask detailed questions about individual-level characteristics (e.g., demographics and personality), which might explain how people respond to foreign-language content. We address this point in the next study.

\newpage
\section[Study II]{STUDY II}
The results of the first study lead us back to the literature to identify what individual-level characteristics might explain who benefits from foreign-language content.

The most relevant literature we identified came from management science where scholars have studied what factors predict successful overseas placements for employees at multinational companies~\cite{arthur1995,caligiuri2000big5,caligiuri2000}. In addition to practical steps such as orientation programs, personality has been found to be important in explaining which employees terminate their overseas placements early~\cite{caligiuri2000big5}. The Big Five personality dimensions are broad and form a good starting place for researching and theorizing the role of personality~\cite{john1999}. All five dimensions 
have been linked to successful overseas placements~\cite{caligiuri2000big5} with particular importance ascribed to 
extraversion and openness~\cite{arthur1995,caligiuri2000}.

Of these, extraversion is thought to be important in helping expatriates to make friends in the new culture and thus contribute to understanding and resilience. In contrast, openness is thought to be more directly related to the attitude of the expatriate toward new experiences, and thus openness should be the most relevant personality dimension in this experiment. Based on this literature, it was hypothesized that individuals with a high level of openness would be more willing to engage with foreign-language content.
Beyond standard measures of openness, 
Caligiuri~\cite{caligiuri2000abos} developed a more detailed scale of attitudinal and behavioral openness building on the work of Budner~\cite{budner1962} examining tolerance of ambiguity. This scale was designed specifically to better understand the role of openness in helping expatriates adapt to new cultures.

\subsection{Method and approach}
We follow the same approach as Study I, but added additional questions drawn from the literature to examine personality. We specifically investigated the Big Five using the Ten Item Personality Inventory~\cite{gosling2003} in addition to a separate, more in-depth measure of attitudinal openness from Caligiuri~\cite{caligiuri2000abos}.

We also introduced a third, ``pretranslation'' condition into our between subjects comparison. This condition showed six English reviews, but stated that three of the reviews had been translated from Spanish. It further offered a button to ``See original'' that, if clicked, showed the Spanish translation of the review. We randomly assigned subjects between the no translation, translation, and pretranslational conditions. We set the proportions to be 20\%, 40\% and 40\% for each condition respectively in order to have more subjects in the translation and pretranslation conditions.

We further changed crowdsourcing platforms and ran this second studying using Prolific.ac. Prolific is designed specifically for academic studies and a number of demographic variables are available about subjects without requiring these questions to be asked in the study itself. This allowed us to keep our experiment short while still investigating demographic variables in-depth. In particular, we examined the following variables from Prolific: gender, age, highest education level, religious affiliation, political affiliation, and Caucasian/non-Caucasian ethnicity. We added to these the Big Five personality characteristics and the attitudinal openness scale.

Although Prolific has a large international pool of workers, we restricted responses to people in the United States in order to test/replicate the findings of Study I. We also required Prolific subjects to have a 90\% or higher acceptance rate (the highest rate that can be required). Payments on Prolific are dominated in British Pounds, and we paid our subjects \textsterling 0.60 GBP (approximately \$0.80 USD at the time of the experiment) for their time. This payment is higher than Study I due to the additional time required to answer the personality questions.

\subsection{Results}
300 subjects completed this second experiment, but we discarded 7 subjects who answered attention check questions incorrectly, 28 subjects who suspected the study was about language, and two subjects who had Internet connectivity issues. Once again most subjects had either no Spanish proficiency (N=125) or basic proficiency (N=113). We discarded 25 subjects with intermediate or higher levels of Spanish and two subjects who were not native speakers of English. We analyzed the remaining 236 subjects.

Per our randomization, 40 subjects saw the en3es3 condition with no translation option, 92 saw this condition with reviews in Spanish and the option to translate to English, and 104 saw this condition with reviews in English (and were told they had been translated from Spanish).

Our Prolific subjects were all born in the United States and native speakers of English. They ranged in age from 18 to 73 (mean 34), and a slightly larger proportion were male than female (58\% male). Most (79\%, 184 subjects) had Caucasian ethnicity, 14 had African ethnicity, 10 had East Asian ethnicity, nine of mixed ethnicity, and the remainder were of other ethnicities. Statistical tests showed similar age, gender, and Caucasian/non-Caucasian distributions across the three treatment conditions.

\subsubsection{Translation conditions}
\begin{figure*}
\includegraphics[width=\textwidth]{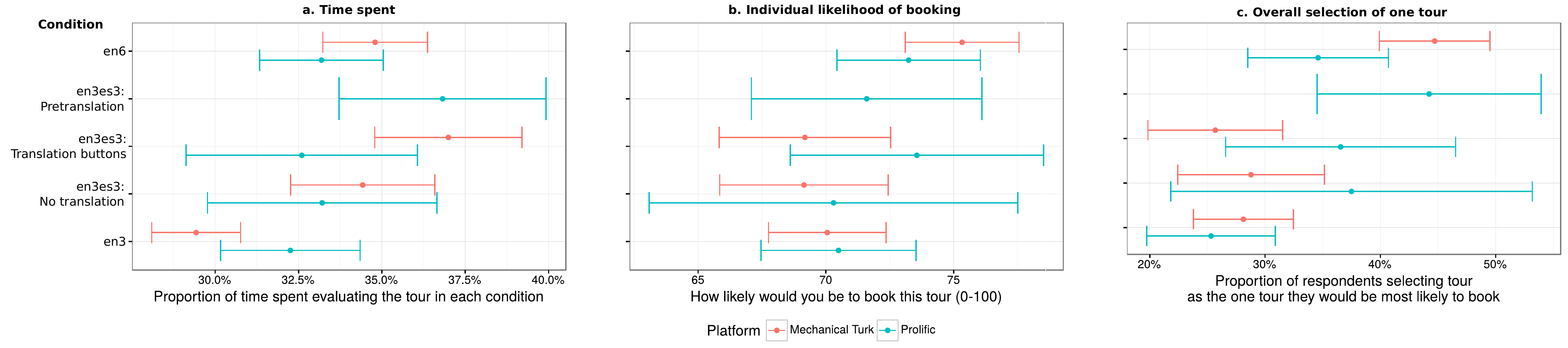}%
\caption{Mean values with 95\% confidence intervals. There was wide variance in the (a) time spent, (b) individual ratings of each tour, and (c) overall selection of one tour.}
\label{fig:prolific_outcomes}
\end{figure*}

The data from the Prolific subject pool showed greater similarity between the three conditions than did the data from Mechanical Turk subjects.
Subjects spent a larger proportion of time in en6 than en3 (34\% vs.\ 33\%, significant only with controls $p=0.03$) while the proportion of time spent in en3es3 (33\%) was not significantly different from the other conditions (Figure~\ref{fig:prolific_outcomes}a). 

Due to the greater similarity and smaller sample size, there were no statistically significant differences between the individual booking likelihoods or the overall booking choices.
The booking likelihoods in the condition with reviews pretranslated showed high variance (sd 23.2), and the mean value was not significantly different from that of the en6 condition (71.6 vs.~73.2, $p>0.5$). Neither of these values differed significantly from the en3es3 condition with the option to see translations (mean 73.6, $p>0.5$, Figure~\ref{fig:prolific_outcomes}b).

Importantly, however, the key observation from Study I was replicated: \emph{subjects who clicked to see at least one translation of a Spanish review rated the tour in the en3es3 condition more highly than the subjects who did not click to see any translations}.

A similar proportion of subjects (26\%) clicked to view one or more translations when given the option. The subjects who clicked to see translations indicated they were more likely to book the en3es3 tour than those who did not click to see any translations.
Subjects not clicking a translation button rated the tour a mean 70 points while those clicking to see one or more translations rated the tour a mean 84 points. This difference was statistically significant with both a t-test ($p=0.007$) and multivariate regression with control variables ($p=0.01$).

In fact, subjects who clicked to see at least one translation in the en3es3 also rated the en3es3 tour more highly than the en6 tour (mean ratings 84 and 73, $p=0.01$ for a t-test, $p=0.03$ for a multivariate regression with control variables).
In contrast, there was no significant difference between the ratings of the en3es3 tour for those not clicking to see translations and the en6 condition (mean ratings 70 and 73, $p=0.35$ for a t-test, $p=0.13$ with control variables).
The choice of one tour as the most preferred tour showed a similar pattern, but none of the differences were statistically significant.

\subsubsection{Individual differences}
We now seek to understand what individual-level characteristics might explain which subjects found the Spanish reviews more useful as well as which subjects clicked the translation buttons. For this subsection we restricted our analysis to the 92 subjects in the translation condition.

As mentioned, subjects who clicked to see one or more translations rated the en3es3 tour significantly higher than those who did not click to see any translations. Surprisingly, however, no demographic or personality variables were associated with higher ratings of the tours. 

Given that those who viewed translations and those who did not behaved so differently, we turned to analyze which subjects clicked to see one or more translations.
Contrary to expectations again, however, no personality or demographic variables were significantly associated with clicking to see translations.

Perhaps unsurprisingly there was a strong relationship between the importance given to reviews and clicking to see one or more translations (84 vs.~94 points, $p=0.001$).
The direction of this relationship could go either way: people who clicked on the translation buttons could have found the reviews more helpful or, alternatively, people who feel reviews are very important in general could have clicked to see the translations.
More notably, the mean importance given to reviews in all three conditions was significantly higher among subjects clicking to see translations (86 vs.~94 points, $p<0.001$). This suggests the latter interpretation is more likely: the people most likely to click to see translations are people who place large importance on user reviews.

\subsection{Discussions}
Study II reproduced the key result of Study I using a different pool of subjects. Namely, \emph{subjects who clicked to see a translation of a Spanish review rated the tour in the en3es3 condition more highly than did the subjects who did not click to see any translations}. We did not ask individual demographic or personality information from Mechanical Turk subjects, but in general both subject pools are generally more educated than the US population as a whole~\cite{ross2010}. While Prolific respondents in this study were slightly more likely to be male, US-based users of Mechanical Turk are, in general, more likely female.\footnote{Ross et al.~\cite{ross2010} found US-based Mechanical Turk workers were 37\% male and 55\% had a Bachelor's degree or higher.}

Despite the additional demographic and personality variables analyzed in Study II, the understanding produced of who benefits most from foreign-language content is incomplete. The action that most distinguished participations was whether they clicked to see translations. Those who did rated the tour most highly. None of the personality variables drawn from the literature had a significant effect, but the subjects clicking to see translations were those subjects who placed large importance on user reviews.

Pretranslating user-generated content to the user's language might seem like a way for more people to benefit from foreign-language content. However, the results showed high variance in this condition. As in Study I, there were likely subjects who were influenced simply by the presence of foreign-language content. The translation buttons provide a simple way of identifying those who find the foreign-language content most useful. More generally, the results indicate that platform operators should consider tracking all types of interaction with foreign-language content and personalizing the display of that content on a per-user basis. Some people may not wish to see this content, others will want to see it but need the option of seeing translations, and still others will want to read the content in its original languages. This study concentrated on people with no or limited proficiency in Spanish as a starting point, but people with higher foreign-language proficiencies are an obvious next step. A large proportion of the human population is bilingual and the needs of these users must be considered~\cite{grosjean2010,steichen2014}. 

\section[Conclusions]{CONCLUSIONS}

The increasing amount of content in different languages online has resulted in large differences in the information available in different languages. Within this article, we have specifically focused on user reviews online, but the findings are also relevant to how foreign-language content is presented on social media and other platforms. Our experimental findings indicate the helpfulness that people derive from foreign-language reviews is a complex topic.

At the broadest, most general level, displaying foreign-language reviews is largely neutral. Our subjects in both studies found having three English and three Spanish reviews very similar to having only three English reviews, which was contrary to our hypothesis that more reviews (even in a foreign language) would be more helpful (H\ref{h:number}).
At the same time, displaying foreign-language content, even in the absence of translations, did not create the overall negative effects that some interface designers fear (and hence seek to separate content in different languages).

The finding in Study I that the tour with six English reviews was most preferred confirms the second hypothesis that consumers most value reviews in their first language (H\ref{h:first}). In general this finding would suggest interfaces should prioritize reviews in a user's first language over other reviews. However, our interface did not provide any additional information about the reviews such as their dates/recency or about the reviewers such as their locations. Future work is needed to understand how such factors intersect with language preferences. 

The context of the product being evaluated is also an area for future work. London is in an English-speaking country and therefore our subjects might have been less expecting or tolerant of non-English reviews (despite findings that a quarter of all London reviews on TripAdvisor are non-English~\cite{hale2016tripadvisor}). In future work, we would like to repeat the experiment with a non-English destination. For example, the same experimental setup and conditions could be used with bicycle tours in Madrid rather than London. In such a case, having Spanish as a local language might result in Spanish-language reviews being more expected, tolerated, or even considered to have prestige or trustworthiness as local content~\cite{sen2015barriers} and thereby render different experimental results. Future work should also consider different products beyond tours or tourist attractions. In particular, the fact that a tour is experienced in a group setting may result in foreign-language reviews being viewed differently than when a good is individually consumed, such as a hotel room.

Native English speakers display the lowest levels of bilingualism online \cite{eurobarometer2011,hale2014twitter,hale2014wiki} and native English speakers were selected as a starting point for this research precisely so that the findings here likely provide a lower-bound on the amount of helpfulness people derive from content in other languages. It is expected that native speakers of smaller-sized languages will be more tolerant of and derive more value from foreign-language reviews (in line with findings that speakers of smaller-sized languages show higher levels of bilingualism online~\cite{hale2014wiki}). This is something that future work should examine empirically. 
Foreign-language content can be especially important for languages with fewer speakers online and/or where many speakers access the Internet from mobile devices and may find it more difficult to contribute longer-form content. The successful use of foreign-language content could help overcome the initial ``cold start'' problems of acquiring users in a new or smaller-sized language on a user-generated content platform if the foreign-language content can be used to fill gaps in content.

The most important outcome of both studies relates to subjects' use of the option to see translations and the different and opposite value of foreign-language reviews for subjects who clicked versus subjects who did not click to see translations. Overall, 26--28\% of subjects given the option to see translations of the Spanish-language reviews clicked to view at least one translation. For these subjects, the Spanish-language reviews were positively received, and the subjects were more likely to indicate they would book the tour with these reviews. In contrast, subjects who did not click to see any translations despite being given the option viewed the tour more negatively and were less likely to book the tour with Spanish-language reviews.
This behavior was consistent in both Study I using Mechanical Turk and Study II using Prolific.

Although it is possible that the addition of the translation buttons themselves created this effect by drawing more attention to the fact that the reviews were not in English---something that would need to be tested with an eye tracking study---it is likely that the use of the translation buttons simply clearly separated subjects with different pre-existing attitudes toward foreign-language reviews. That is to say that the control group contained subjects with a mixture of attitudes toward foreign-language reviews but it was not possible to observe these attitudes. The introduction of the translation buttons could simply have created the opportunity for subjects to differentiate themselves according to their attitudes toward foreign-language reviews: those with pre-existing positive attitudes could have been more likely to click the translation buttons while those with pre-existing negative attitudes could have been less likely to click the translations buttons and thus produce (at least some of) the experimental effects observed.

In Study II we tested individual-level characteristics that could explain this behavior and the hypothesized pre-existing attitudes. Openness was expected to be a key individual trait, but in fact was found not to play a significant role. Indeed, none of the personality variables examined in Study II were significant.
The largest predictor of using the translation buttons was a high importance on reviews. 

The finding that the mere presence of foreign-language content can alter perceptions fits with branding and country-of-origin studies for products in marketing research \cite{alsulaiti1998,klein2002,leclerc1994}. For example, Leclerc et al.~\cite{leclerc1994} found that when fictitious company names were pronounced in French rather than English, the attitudes of US subjects toward the fictitious brand and the hedonism of its products changed.

Preliminary analysis of the comments left in the optional free text questions of Study I support the idea that the mere presence of foreign-language reviews had an impact for a small number of subjects. Some subjects saw the foreign-language reviews as a positive, ``because it indicates the company is flexible and multinational,'' while other subjects saw the foreign-reviews negatively. In the most negative comment one subject stated ``\dots{}I get enough Spanish here in the US. I'd rather not be around Spanish speaking people when I visit London.'' The majority of comments left within the en3es3 condition mentioned the reviews positively without any reference to their language or made a neutral statement such as ``last couple of reviews were in a foreign language.''

This study assessed the overall effects of foreign-language content and the impact of including a translation option.
It took a first step toward analyzing how individual level characteristics (demographics, personality, etc.) affect people's attitudes toward foreign-language content, but the results were not fully satisfactory and this is an important area of investigation that is severely lacking in academic research. 
Perhaps the best recommendation for platform designers and operators at this point is to personalize the showing of foreign-language content and translations. If users interact with foreign-language content and/or click to view translations, more foreign-language content and translations should be shown to these users if available. However, if users ignore the foreign-language content and translation buttons they are offered, it might be appropriate to hide further foreign-language content from these users.
Special attention should be paid, however, to bilingual users as these users are likely to appreciate content in their multiple languages.

Future work should also explore alternative designs and evaluation measures such as adding ``this review is helpful'' buttons and/or having reviews of different polarities in different languages (e.g., positive reviews in Spanish but negative in English or vice versa). The roles that cultural (dis)similarity or affinity play are also unclear. 

This study, as a first step in assessing the impact of including foreign-language reviews, has indicated that reactions to and the helpfulness derived from foreign-language reviews is a complex topic. We hope that this study will inspire future work on this important topic. As the quantity of reviews and other content in all languages increases, much further research is needed in order to understand when to use and how to present foreign-language content in different contexts to users with different linguistic backgrounds and attitudes.

\newpage
\section[Acknowledgments]{ACKNOWLEDGMENTS}
We would like to thank our academic colleagues, friends at Meedan, and anonymous reviewers who provided helpful feedback on this research. This publication was supported by the John Fell Oxford University Press (OUP) Research Fund, The Alan Turing Institute under the EPSRC grant EP/N510129/1, the University of Oxford's Economic and Social Research Council (ESRC) Impact Acceleration Account and Higher Education Innovation Fund (HEIF) allocation, and the European Union's Horizon 2020 research and innovation programme under the Marie Sklodowska-Curie grant agreement No.\ 656439.

\balance{}

\bibliographystyle{SIGCHI-Reference-Format}

\appendix
\section[Question wordings]{QUESTION WORDINGS}

\begin{itemize}
	\item Questions asked of each tour:

	\begin{itemize}
		\item What type of tour is this? Walking, Bus, Bike/Cycle (random order)
		\item If you were to book a tour of this type, how likely would you be to book this particular tour? Visual analogue scale with end points ``Very unlikely'' and ``Very likely'' mapped to integers 0--100
		\item In reaching this decision how important were the following factors: Title, Photo, Description, Visitor reviews. Visual analogue scales with end points ``Very unimportant'' and ``Very important'' mapped to integers 0--100
		\item Any specific comments? (optional)
	\end{itemize}

	\item Overall selection of one tour
	\begin{itemize}
		\item If you were to book one of the three tours that you have just examined, which tour would you be most likely to book? (You may click on a tour to quickly see its listing again if you like.)
		\item How confident are you in your choice? Visual analogue scale with end points ``Not at all confident'' and ``Very confident'' mapped to integers 0--100
		\item In making the decision to select the tour that you did, how important was each of the following elements? Title, Photo, Description, Visitor reviews. Visual analogue scales with end points ``Very unimportant'' and ``Very important'' mapped to integers 0--100
		\item Do you have any previous experience or knowledge of any of the tours that you examined?
		\begin{itemize}
			\item I have had previous experience with this tour
			\item I have \textbf{not} had previous experience with this tour
		\end{itemize}

	\end{itemize}
	
	\item Big Five (Study II only)\\ 
	All questions were asked using visual analogue scales with end points ``Strongly disagree'' and ``Strongly agree'' mapped to integers 0--100. Question order was randomized for each subject. Details are given in Gosling et al.~\cite{gosling2003}
	
	\item Openness (Study II only)\\
	All questions were asked using visual analogue scales with end points ``Strongly disagree'' and ``Strongly agree'' mapped to integers 0--100. Question order was randomized for each subject.
	\begin{itemize}
		\item I like being with unpredictable people
		\item I like parties where I know most of the people more than ones where all or most of the people are complete strangers
		\item What we are used to is always preferable to what is unfamiliar 
		\item I would like to live in a foreign country for a while
		\item A person who leads an even, regular life in which few surprises or unexpected happenings arise, really has a lot to be grateful for
		\item Other cultures fascinate me
		\item I would prefer a foreign movie be subtitled rather than dubbed
		\item Foreign language skills should be taught in (as early as) elementary school
	\end{itemize}

	\item Language ability questions
	\begin{itemize}
		\item Prior to reading the information above, did you suspect that this study was about foreign-language reviews? Yes, No.
		\item How would you rate your ability to read English?
		\begin{itemize}
			\item I do not read English at all
			\item I read English at a basic or beginning level
			\item I read English at an intermediate level
			\item I read English at an advanced level
			\item English is my native language
		\end{itemize}
		\item How would you rate your ability to read Spanish? (same options as for English)
		\item Foreign experience (Study II only)\\
		Question order randomized; Answer choices: Yes, No, Prefer not to say
		\begin{itemize}
			\item I am fluent in a language besides English
			\item I have spent time living in another country
			\item I have moved or been relocated a substantial distance (e.g., state to state or overseas)
			\item I have studied a foreign language
		\end{itemize}

		\item Is there anything else you would like to share with us? (optional)
	\end{itemize}

\end{itemize}

\end{document}